\documentclass[final,3p,12pt]{elsarticle}
\usepackage{graphicx,amsbsy,amsmath,epsfig,natbib,float}
\usepackage{amssymb,latexsym,wasysym,pifont,mathrsfs}
\usepackage{comment,verbatim,multirow,array,sidecap,color}
\usepackage{lineno,setspace,geometry,soul,cancel}
\usepackage{mciteplus}
\newfloat{widefig}{thp}{lop}
\journal{Physica A}

\begin{document}

\begin{frontmatter}
\title{Critical properties of deterministic and stochastic sandpile
  models on two-dimensional percolation backbone}

\author{Himangsu Bhaumik} \author{S. B. Santra}
\ead{santra@iitg.ernet.in}

\address{Department of Physics, Indian Institute of Technology
  Guwahati, Guwahati-781039, Assam, India.}

\date{\today}

\begin{abstract}
  Both the deterministic and stochastic sandpile models are studied on
  the percolation backbone, a random fractal, generated on a square
  lattice in $2$-dimensions. In spite of the underline random
  structure of the backbone, the deterministic Bak Tang Wiesenfeld
  (BTW) model preserves its positive time auto-correlation and
  multifractal behaviour due to its complete toppling balance, whereas
  the critical properties of the stochastic sandpile model (SSM) still
  exhibits finite size scaling (FSS) as it exhibits on the regular
  lattices. Analyzing the topography of the avalanches, various
  scaling relations are developed. While for the SSM, the extended set
  of critical exponents obtained is found to obey various the scaling
  relation in terms of the fractal dimension $d_f^B$ of the backbone,
  whereas the deterministic BTW model, on the other hand, does not. As
  the critical exponents of the SSM defined on the backbone are
  related to $d_f^B$, the backbone fractal dimension, they are found
  to be entirely different from those of the SSM defined on the
  regular lattice as well as on other deterministic fractals. The SSM
  on the percolation backbone is found to obey FSS but belongs to a
  new stochastic universality class.
\end{abstract}


\end{frontmatter}

\section{Introduction} 
The term ``Fractal'' was coined by B. Mandelbrot \cite{mandelbrot77}
in order to address the notion of naturally occurring self-similar
structures. At the same time to model the fractal objects, the concept
of lattice with non-integer dimension (fractal lattice) was also
introduced in the same spirit \cite{dharJMP77}. Various lattice
statistical models are well studied on such fractal objects not only
to investigate how the well-established theories like
$\epsilon$-expansion or real space renormalization group work on such
objects but also to verify the effect of non-integer dimension on the
scaling relations which are derived straightforwardly on a hypercubic
lattice. For example, study of critical phenomena on deterministic
fractal lattices through renormalization-group technique
\cite{gefenPRL80} as well as numerical analysis
\cite{carmonaPRB98,*pruessnerPRB01,*windusPHYA09} reveals that the
critical properties of several lattice statistical models are affected
not only by the non-integer dimension but also by various other
topological aspects of the fractal lattice (such as ramification,
connectivity or lacunarity). Backbone of the incipient infinite
percolation cluster \cite{stauffer,*bunde-havlin} in two-dimensions
($2$D) at the percolation threshold is a random fractal lattice whose
various scaling properties are well known
\cite{hongPRB84,grassbergerPHYA99,*barthelemyPRE99,bundePRE97}.
Several dynamical models are also studied on percolation cluster for
its wide application, such as random walk on percolation cluster
\cite{rammalJPA83}, flow in porous media
\cite{grovaJPCS11,najafiPHYA15}, absorbing state phase transition
\cite{leePRE13}, etc.  In all such studies, the emergence of
non-trivial results occurs due to the coupling of the fractal nature of
the underlying object to the model's critical dynamics.

On the other hand, the concept of Self Organized Criticality (SOC) was
introduced by Bak, Tang and Weisenfeld (BTW) in order to understand
the spontaneous emergence of spatial and temporal correlation (and
hence the criticality) of a wide class of slowly driven natural
systems \cite{bak,*jensen}. BTW sandpile model then becomes a generic
model to study the SOC \cite{btwPRL87,*btwPRA88}. Several variants of
the BTW model have been extensively studied on regular lattices and
many analytical, as well as numerical results exist in the literature
\cite{dharPHYA99a,*dharPHYA06}. Among them, the Stochastic Sandpile
Model (SSM) \cite{dharPHYA99a,*dharPHYA06} is a well-studied model for
its clean scaling behaviour which does not exist in BTW model. It is
widely accepted that the BTW model has a multiscaling behaviour
\cite{deMenechPRE98,tebaldiPRL99} due to its complete toppling balance
\cite{karmakarPRL05} and positive auto-correlation in avalanche wave
series \cite{deMenechPRE00,deMenechPHYA02}, whereas the SSM does not
show such correlation and consequently follows finite size scaling
(FSS) ansatz. Recent numerical studies of the SSM have been carried
out not only on various regular lattices of integer dimension but also
on various kind of deterministic fractal lattices
\cite{huynhJSM11,huynhPRE10,huynhPRE12} and the results confirm the
existence of robust FSS behaviour of the SSM across different regular,
as well as fractal lattices though the universality class depends on
the space or fractal dimension of these lattices. The fractal lattices
considered for such studies were deterministic, the properties of SSM
as well as BTW on random fractal lattices are yet to be studied. It is
then intriguing to study both the BTW and the SSM on the percolation
backbone and investigate whether the random fractal structure of the
backbone can destroy the positive auto-correlation of BTW model and
the model would belong to a new universality class, whether the SSM
can still preserve its robust FSS behaviour and exhibits behaviour of
a new stochastic universality class. In this article, both BTW and SSM
are studied on the percolation backbone generated on the square
lattice in $2$D and their scaling behaviour estimating an extended set
of exponents through extensive numerical simulations are reported.
 
\section{The models}
Infinite percolation networks are obtained generating percolation
clusters on the $2$D square lattice employing the well-known
Hoshen-Kopelman algorithm \cite{hoshenPRB76} with the open boundary
condition. A backbone network is then extracted from an infinite
cluster at the percolation threshold ($p_c=0.59278$) using irreducible
configuration of articulation sites (if removing a site breaks the
cluster into two or more parts, then the site is called articulation
site). The details of the algorithm can be found in
Ref. \cite{yinPHYB00}. It has been verified that the fractal dimension
of the backbone is found to be $d_f^B=1.64$ as reported in
\cite{bundePRE97}.

BTW and SSM are briefly described here on a percolation backbone
generated on a $2$D square lattice of size $L\times L$. Each site of
the backbone is associated with a non-negative integer variable $h$
representing the height of the ``sand column'' at that site. Sand
grains are added one at a time to a randomly chosen backbone site and
the height of the sand column of the respective site is increased as
$h_i \rightarrow h_i + 1$. The sand column at any arbitrary backbone
site $i$ becomes unstable or active when its height $h_i$ exceeds a
prefixed critical value $h_c$ and a burst of a toppling activity
occurs by collapsing the sand column distributing the sand grains to
the available nearest-neighbour (nn) sites on the backbone by some
specific rule.  As a result, some of the nn sites may become
upper-critical and lead to further toppling activities in a cascading
manner. Consequently, these toppling activities will lead to an
avalanche. During an avalanche, no sand grain is added and the
propagation of an avalanche stops if all sites of the backbone become
under-critical.

In BTW, the critical height is taken as $h_c = d_i$, where $d_i$ is
the number of available nearest neighbour sites of the $i$th site on
the backbone. The toppling rule of BTW is given by
\begin{equation}
\label{btwr}
h_i \rightarrow h_i - d_i \ \ \ \rm{and} \ \ \ h_j \rightarrow h_j +
1,
\end{equation}
where $j = 1, \cdots, d_i$. Whereas in SSM, the critical height is
fixed as $h_c = 2$ for all the backbone sites. The toppling rule of
SSM is given by
\begin{equation}
\label{ssmr}
h_i \rightarrow h_i - 2 \ \ \ \rm{and} \ \ \  h_j \rightarrow h_j + 1,
\end{equation}
where $j= j1, j2$ are two randomly and independently selected nn sites
out of the $d_i$ nn sites of the $i$th site on the backbone.

Since the backbone is extracted from incipient infinite percolation
cluster, the backbone must consist of the lattice boundary sites. As
both the sandplie models are studied with the open boundary condition,
dissipation of sand grains occurs due to toppling activity on the
lattice boundary sites those belong to the backbone. During
dissipation one sand grain dissipates from the system.

\section{Numerical simulations}
Defining the model on a percolation backbone, repetitive addition of
sand grains drive the system to a steady state that corresponds to the
equal current of incoming flux to the outgoing flux of the sand
grains. Such a situation is identified by the constant average height
of the sand columns. To study the critical behavior of the steady
state of sandpile models on the percolation backbone, different
avalanche properties such as the toppling size $s$, area $a$, and
lifetime $t$ of the avalanches are measured at the steady state. The
toppling size $s$ is defined as the total number of topplings which
occurs in an avalanche, the avalanche area $a$ is equal to the number
of distinct sites toppled in an avalanche, and the lifetime $t$ of an
avalanche is the number of parallel updates to make the unstable
configuration to a stable one where all the sites have $h_i<h_c$. The
system size $L$ is varied from $L=64$ to $L=1024$ in multiples of
$2$. For a fixed system size $L$, $1024$ backbone configurations are
generated. On each backbone, after attaining the steady state for the
considered model, $10^6$ avalanches are neglected and the next $10^5$
avalanches are collected for measurement. Therefore, total $1024\times
10^5$ avalanches are taken for data averaging for a given model with
specific $L$.

\section{Avalanche Morphology}
\begin{figure}[t]
\centerline{\hfill 
  \psfig{file=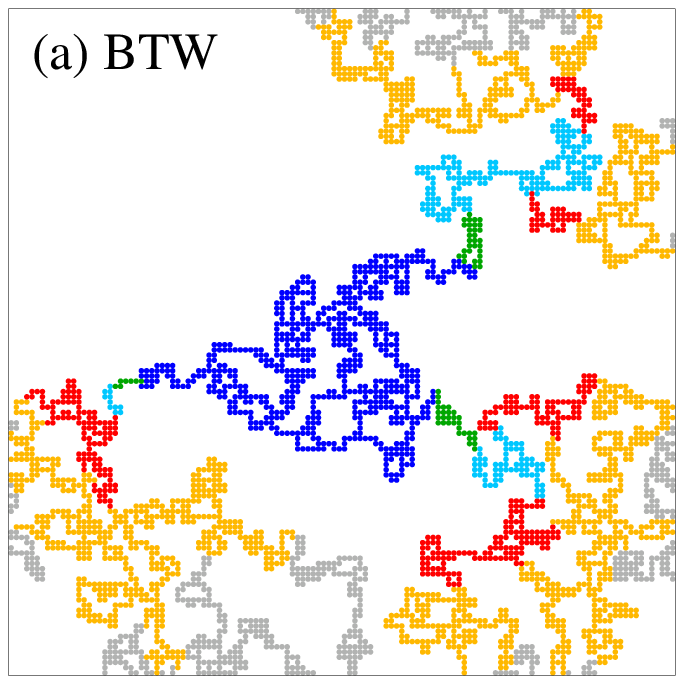,width=0.35\textwidth}\hfill 
 \psfig{file=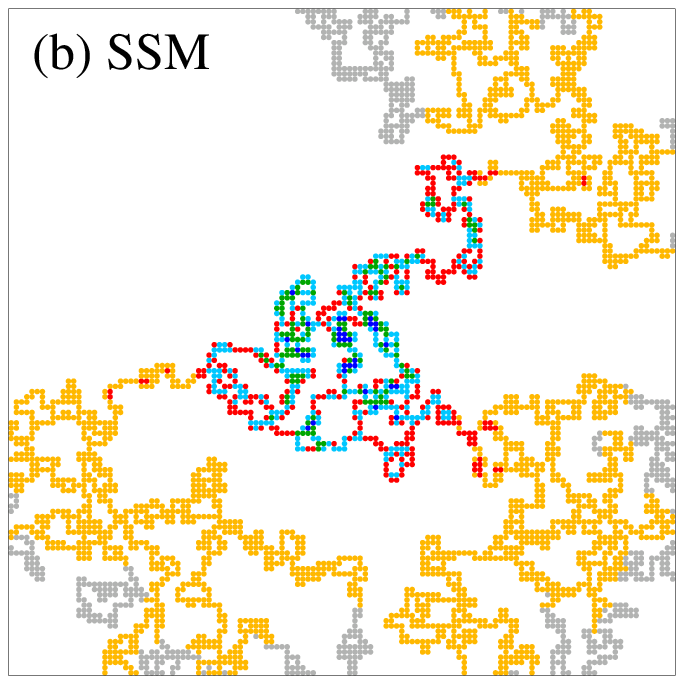,width=0.35\textwidth}
  \hfill}
\caption{\label{morpho}(Colour online) Morphology of a typical
  avalanche cluster of (a) the BTW and (b) the SSM on percolation
  backbone generated on an $L=128$ square lattice. Toppling numbers
  are binned into $5$ equal bins. Different colors correspond to
  different bins: blue, green, cyan, red, and yellow with decreasing
  order of toppling numbers. The gray sites are the site of backbone
  with no toppling. The black border represents the lattice boundary.}
\end{figure} 
The morphology of avalanches in the steady state of both the models
are presented here. Typical large avalanches of BTW and SSM obtained
in their respective steady states are shown in Fig. \ref{morpho}(a)
and \ref{morpho}(b) respectively. These avalanches are obtained on the
same percolation backbone generated on a lattice of size $L=128$,
dropping sand grain at the central part of the backbone (near to the
center of the lattice). The backbone considered here has $3665$ number
of lattice sites. For both the cases, the area of the avalanche is
$80\%$ of the total number of backbone sites. Maximum toppling for BTW
cluster is $154$ whereas that for the SSM is quite high which is
$813$. For both the cases, the toppling numbers are binned into $5$
equal sizes. Different colours correspond to the different bin of
toppling numbers. Blue colour corresponds to the bin of highest
toppling numbers. Green, cyan, red, yellow colours correspond to the
bins of the lower and lower toppling numbers respectively. The gray
color corresponds to the sites of no toppling. It can be seen that the
avalanche in BTW has structured toppling zones, similar to that when
the model was studied on the regular lattice
\cite{christensenPRE93,grassbergerJPF90,*mannaJSP90}. On the other
hand, the avalanche of SSM exhibits random mixing of colors
representing different toppling numbers as that of an avalanche of SSM
on $2$D square lattice \cite{santraPRE07}. It could also be noted here
that though both models preserve the nature of their avalanche
morphology, the maximum toppling or the toppling size is quite larger
than that of the $2$D regular lattice. This is because the walk
dimension is quite high on the backbone than on the regular lattice
(which will be discussed in details in terms of avalanche exponents in
the following section). Sand grains need more steps to travel to the
boundary of the backbone than on a regular lattice starting from the
same point. As a result, more toppling occurs in an avalanche on a
backbone than on a regular lattice.

\section{Multifractal analysis}
\subsection{Probability distribution function}
\begin{figure}[t]
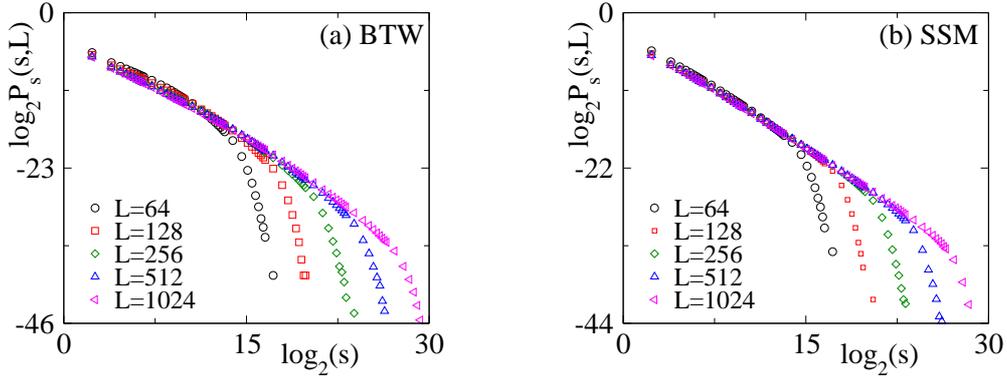

\centerline{\hfill
  \psfig{file=Ps_all_L_BTW.eps,width=0.35\textwidth}\hfill
  \psfig{file=Ps_all_L.eps,width=0.35\textwidth} \hfill}
\caption{\label{ps}(Colour online) Probability distribution of $s$
  for various system sizes $L$ are plotted in (a) for BTW and in (b)
  for the SSM.}
\end{figure} 
The probability distribution of various avalanche properties $x \in
\{s,a,t\}$ are analyzed to characterize the critical steady state of
sandpile models. At the steady state, the probability distribution
function $P_x(x,L)$ of a property $x$ of an avalanche on a percolation
backbone generated on a lattice of size $L$ is expected to obey
power-law scaling as
\begin{equation}
\label{pdf}
P_x(x,L) = x^{-\tau_x}f_x(x/L^{D_x}),
\end{equation}
where $x\in \{s,a,t\}$, $\tau_x$ is the corresponding critical
exponent, $D_x$ is the capacity dimension and $f_x$ is the
corresponding scaling function. Data for toppling size only are shown
in Fig. \ref{ps}(a) and \ref{ps}(b) for BTW and SSM respectively for
various system size $L$.

\begin{figure}[t]
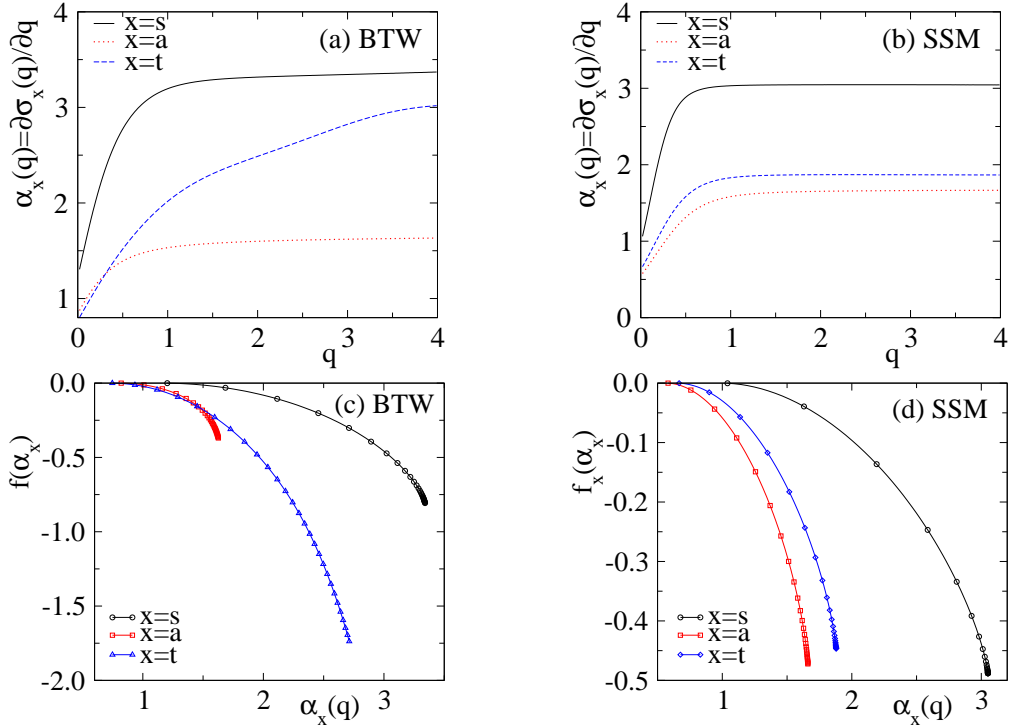

\centerline{\hfill 
  \psfig{file=del_sigma_sat_BTW.eps,width=0.35\textwidth}\hfill 
  \psfig{file=del_sigma_sat_SSM.eps,width=0.35\textwidth}\hfill 
  }
\centerline{\hfill 
  \psfig{file=f_alfa_sat_BTW_q3.eps,width=0.35\textwidth}\hfill 
  \psfig{file=f_alfa_sat_SSM_q3.eps,width=0.35\textwidth}\hfill 
  }
\caption{\label{falfa}(Colour online) (a) Plot of $\alpha_x(q)$
  against $q$ for $x=s,a,t$ for (a) the BTW and (b) the SSM. Plot of
  $\sf f_x(\alpha_x)$ against $\alpha_x(q)$ for BTW and SSM in (c) and
  (d) respectively. Accumulation of points are observed clearly for
  each $x\in \{s,a,t\}$ in the case of SSM.}
\end{figure}
To estimate the values of the exponents $\tau_x$ and $D_x$ defined in
Eq. (\ref{pdf}), the concept of moment analysis
\cite{karmakarPRL05,lubeckPRE00a} for the various avalanche properties
has been employed. The $q$th moment of $x$ is then given by
\begin{equation}
\label{qmnts1}
\langle x^q \rangle = \int_0^{x_{max}} x^{q}P_x(x,L)dx \sim
L^{\sigma_x(q)}
\end{equation}
and $\sigma_x(q)=[q+1-\tau_x]D_x$. If the probability distributions
obey the scaling form given in Eq.\ref{pdf}, the moment scaling
function $\sigma_x(q)$ would be a piece wise linear : $\sigma_x(q)=0$
for $q<\tau_x-1$ and $\sigma_x(q)=D_x[q+1-\tau_x]$ for $q>\tau_x-1$
for $x\in \{s,a,t\}$. Hence, a multiscaling analysis
\cite{deMenechPRE98,tebaldiPRL99,stellaPHYA01} will be useful in which
a spectra of singularity strengths ${\sf f}_x(\alpha_x)$ are
obtained. The singularity strengths ${\sf f}_x(\alpha_x)$ can be
obtained by Legendre transformation of $\sigma_x(q)$ as
\begin{subequations}
\begin{equation}
 {\sf f}_x\left [ \alpha_x(q)\right ]-\sigma_x(q)=- q\alpha_x(q)
\end{equation}
\mbox{and}
\begin{equation}
 \alpha_x(q)=\partial \sigma_x(q)/\partial q
\end{equation}
\end{subequations}
which are expected to converge at higher values of $q$. Therefore, a
plot of ${\sf f}_x(\alpha_x)$ vs $\alpha_x$ should exhibit an
accumulation of points for large $q$ in the ${\sf f}_x$-$\alpha_x$
plane where $\alpha_{x,max}$ and ${\sf f}_{x,min}$ correspond to $D_x$
and $-(\tau_x-1)D_x$ respectively. The analysis not only confirms
whether the models exhibit FSS or not but also estimates the exponents
in $L\rightarrow\infty$ limit.

Following \cite{stellaPHYA01,chhabraPRL89}, the technique of direct
empirical determination of ${\sf f}_x$ has been applied here. Two
quantities $\mbox{log}\langle x^q (L)\rangle/\mbox{log}L$ and $\langle
\mbox{log}(x)x^q(L) \rangle / [\mbox{log}(L)\langle x^q(L) \rangle]$
have been calculated for finite $L$ and then extrapolated to
$L\rightarrow \infty$ limit by a suitable logarithmic correction
proposed by Manna \cite{mannaPHYA91} to estimate $\sigma_x(q)$ and
$\alpha_x(q)$ for both the models. The values of ${\sf f}_x
\left[\alpha_x(q)\right ]$ have been calculated for the moment $q$ in
the range $0<q\le 4$ with $0.01$ interval. For different avalanche
properties $x=s,a,t$, the plots of $\alpha_x(q)$ against $q$ are shown
in Fig. \ref{falfa}(a) for BTW model and in Fig \ref{falfa}(b) for
SSM. It can be seen that all the $\alpha_x(q)$s do not converge for
the BTW model upto $q=4$ whereas in the case of SSM, there is a clear
convergence of $\alpha_x(q)$, which confirms that though the spatial
structure is random fractal the BTW retains its multifractal behavior
whereas the SSM obeys FSS. Different spectra of ${\sf f}_x(\alpha_x)$
are plotted in Figs. \ref{falfa}(c) and \ref{falfa}(d) for BTW and SSM
respectively. For the case of BTW model, the points are not
accumulated at a point on the ${\sf f}-\alpha$ plane which is more
prominent for avalanche time $t$. Hence, BTW model exhibits true
maultifractal behaviour and no exponent is possible to
extract. However, it is recently shown in the $2$D induced model
\cite{najafiJSTAT18}, $2$-dimensional cross-section of site dilutated
cubic percolatioan lattice, the BTW exponents have similarities with
the $2$D Ising universality class \cite{najafiJSTAT15} and satisfy
some hyper-scaling relations. Whereas for SSM, the accumulation of
points at large $q$ corresponds to $(\alpha_{x,max}, {\sf
  f}_{x,min})$. The values of $(\alpha_{x,max}, {\sf f}_{x,min})$ are
found to be ($3.062,-0.472$), ($1.663,-0.479$), and ($1.872,-0.442$)
for $x=s,a$, and $t$ respectively and the values of different critical
exponents of SSM are then estimated as $\tau_s=1.154$, $\tau_a=1.288$,
$\tau_t=1.236$, $D_s=3.062$, $D_a=1.663$, $D_t=1.872$. Note that the
values of the exponents obtained for SSM on the backbone are
completely different from those known on the square lattice and other
deterministic fractals. A detailed comparison of the exponents on the
square lattice and deterministic fractal lattices with those on the
percolation backbone is given in Table \ref{table1}. Thus SSM on the
percolation backbone obeys FSS but belongs to a new stochastic
universality class.
\begin{table}[t]
\centering
\begin{tabular}{lllll}
\hline
\hline
 Exponent & Square Lattice & SSTK & Arrowhead & Backbone\\
& ($d=2$) & ($d_f=1.46$) & ($d_f=1.58$)  & ($d_f^B=1.64$)\\
\hline
$\tau_s$ & 1.273(2) & 1.13(2)   & 1.173(1) & 1.154(19)\\
$\tau_a$ & 1.382(3) & 1.273(11) & 1.298(1) & 1.288(13)\\
$\tau_t$ & 1.489(9) &  1.21(2)  & 1.279(2) & 1.236(14)\\
$D_s$    & 2.750(6) & 2.94(3)   & 2.793(2) & 3.06(2)\\
$D_a$    & 1.995(3) & 1.466(5)  & 1.584(1) & 1.66(1)\\
$D_t$    & 1.532(8) & 1.81(1)   & 1.673(1) & 1.87(1)\\
$\gamma_{sa}$ & 1.23(1)  & --    &--        & 1.784(4)\\
$\gamma_{st}$ & 1.70(1)  & --    &--        & 1.547(4)\\
\hline
\hline 
\end{tabular}
\caption{\label{table1} Comparison of different exponents of SSM
  studied on a regular square lattice, deterministic fractal
  [semi-inverse square triadic Koch (SSTK) lattice with fractal
    dimension $1.46$ and arrowhead fractal lattice with fractal
    dimension $\ln(3)/\ln(2)=1.58$] and on random fractal (percolation
  backbone) with fractal dimension $1.62$. The values of the exponents
  for square lattice, SSTK, and Arrowhead are taken from
  Refs. \cite{benhurPRE96,huynhPRE12}. The numbers in the parentheses
  represent the error in the last digit(s) of the value of the
  exponents. In case of backbone (this work), the errors in $\tau$
  estimated from the propagation error in the expression
  $\tau_x=(1-\sf{f}_{x,min}/\alpha_{x,max})$. Similarly the errors in
  $D$ estimated from the least square fit error during extrapolation.}
\end{table}

For lattices with integer dimension it is already known that the average
toppling size $\langle s \rangle$ is equivalent to the average number
of steps of a random walker on a given lattice before it reaches the
boundary starting from an arbitrary lattice point
\cite{nakanishiPRE97,shiloPRE03}. Thus one could get a relation
\begin{equation}
\label{sdt}
\sigma_s(q=1)=D_s(2-\tau_s)=d_w,
\end{equation}
where $d_w$ is the random walk dimension of the lattice
considered. This relation for SSM was not only verified for integer
dimension \cite{nakanishiPRE97} but also for various deterministic
fractal lattice \cite{huynhPRE10,huynhPRE12}. In the present case of
percolation backbone $\sigma_s(q=1)$ for SSM at $L\rightarrow \infty$
is found to be $2.62 \pm 0.02$ which is in agreement with the value of
$d_w \approx 2.64$ estimated by Hong {\em et al.}  \cite{hongPRB84}
performing exact enumeration of random walks on backbone. Note that
taking the measured values of the exponents the quantity $D_s(2-\tau_s)$
has the value $2.59$ which is again within the error bar of the
measured value of $\sigma_s(1)$. Recently, based on an extensive
numerical study Huynh and Pruessner \cite{huynhPRE10,huynhPRE12}
proposed a relation among $D_s$, $d_w$ and spatial dimension $(d)$ as,
\begin{equation}
\label{dab}
D_s=ad+bd_w
\end{equation}
with $a=0.55$ and $b=0.82$. Taking $d=d_f^B=1.64$ \cite{bundePRE97},
the fractal dimension of the backbone, and $d_w=2.64$
\cite{hongPRB84}, the value of $D_s$ will be $3.066$ which is again
consistent with the measured value. The value of $D_a\approx d_f^B$ is
found for both the BTW and SSM as it is expected.

\begin{figure}[t]
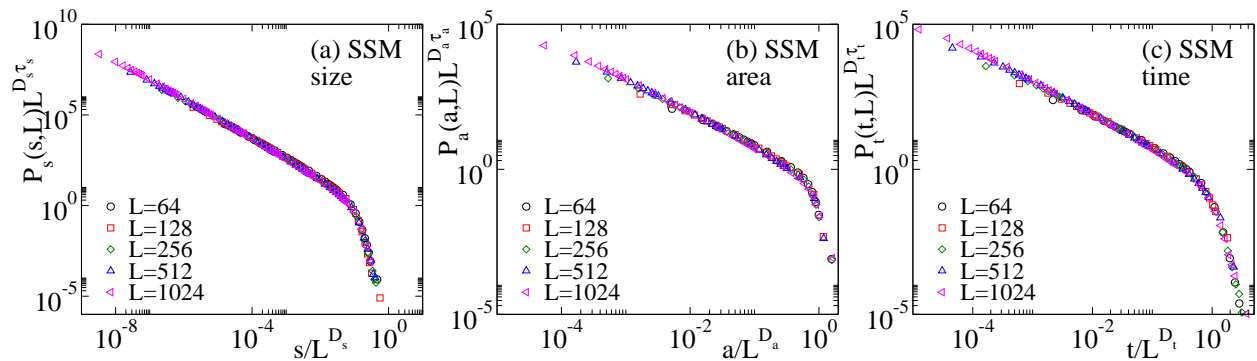

\centerline{\hfill 
  \psfig{file=DC_s.eps,width=0.33\textwidth}\hfill 
  \psfig{file=DC_a.eps,width=0.33\textwidth}\hfill 
  \psfig{file=DC_t.eps,width=0.33\textwidth}\hfill 
  }
\caption{\label{dcpdf}(Colour online) Plot of scaled distribution
  $P_x(x,L)L^{\tau_x D_x}$ of SSM against scaled variable $x/L^{D_x}$
  in (a) for $x=s$, (b) for $x=a$, and in (c) for $x=t$. Different
  curves are for different system sizes $L$. Reasonable collapse of
  data are observed for all the cases.}
\end{figure} 
To verify the measured values of the exponents and the form of the
scaling function defined in Eq. (\ref{pdf}) for SSM, a scaled
distribution $P_x(x,L)L^{\tau_x D_x}$ is plotted against a scaled
variable $x/L^{D_x}$ in Fig. \ref{dcpdf} for different avalanche
properties. For all the cases, the reasonable data collapse confirms
the FSS in the SSM defined on the percolation backbone.

\subsection{Conditional expectation}
\begin{figure}[t]
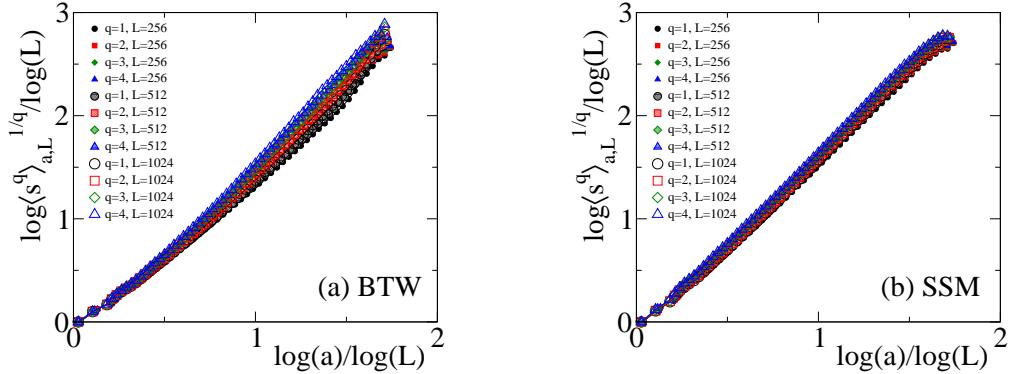

\centerline{\hfill 
  \psfig{file=CE_moment_diffL_BTW.eps,width=0.35\textwidth}\hfill 
  \psfig{file=CE_moment_diffL_SSM.eps,width=0.35\textwidth}\hfill
}
\caption{\label{CEq}(Colour online) Plot of scaled conditional moment
  scaling function $\beta_a\kappa(q)/q=\log(\langle
  s^q\rangle_{a,L}^{1/q})/\log(L)$, against $\beta_a=\log(a)/\log(L)$
  for (a) BTW and (b) SSM for three different system sizes $L=256$,
  $512$, $1024$ and different values of $q$. }
\end{figure}

The critical behaviour of sandpile models on the backbone is further
investigated by studying the conditional expectation values
\cite{christensenJSP91} of the avalanche properties through moment
analysis technique following Refs.
\cite{deMenechPRE98,tebaldiPRL99}. For a fixed system size $L$, the
$q$th moment of the conditional expectation $\langle x^q\rangle_{y,L}$
of a property $x$ keeping another property $y$ fixed at a certain
value, is defined as \cite{christensenJSP91},
\begin{equation}
\label{cpd}
\langle x^q\rangle_{y,L} = \int_{0}^{\infty} x^qP_{x|y,L}(x|y,L)dx
\end{equation}
where $P_{x|y,L}(x|y,L)$ is the conditional probability of property
$x$ for a fixed value of $y$ and a fixed system size $L$. If
$P_x(x,L)$ obeys FSS, $P_{x|y,L}(x|y,L)$ can be assumed as
$P_{x|y,L}(x|y,L)\sim \delta(x-y^{\gamma _{xy}})$ in the
$L\rightarrow\infty$ limit where $\gamma_{xy} $ is a critical
exponent. The quantity $\langle x^q\rangle_{y,L}$ is
then expected to scale with the other property $y$ as
\begin{equation}
\label{cev}
\langle x^q\rangle_{y,L} \sim y^{\kappa(q)}
\end{equation}
where $x\in \{s,a,t\}$ and $\kappa(q)=q\gamma_{xy}$ is a $L$
independent moment exponent. However, for finite system, the quantity
$y$ is expected to scale with the system size $L$ as $y\sim
L^{\beta_y}$ and the conditional moment should be given by
\begin{equation}
\label{conml}
\langle x^q\rangle_{y,L} \sim L^{\beta_y\kappa(q)}
\end{equation}
where $\beta_y\kappa(q)$ would be the conditional moment scaling
function. If the system obeys FSS, the quantity $\beta_y\kappa(q)/q$
will be independent of $q$ and would be equal to $\beta_y \gamma_{xy}$
\cite{tebaldiPRL99}. Thus a plot of $\beta_y\kappa(q)/q$ versus
$\beta_y$ will give a unique slope $\gamma_{xy}$ for various values of
$q$ and $L$. To measure $\beta_y\kappa(q)/q$ the quantity
$\log(\langle x^q \rangle_{y,L}^{1/q})/\log(L)$ is calculated and
plotted against $\beta_y=\log(y)/\log(L)$ for $x=s$ and $y=a$ in
Fig. \ref{CEq}(a) for BTW and in Fig. \ref{CEq}(b) for the SSM for
$q=1,2,3$, and $4$ and for three different system sizes, $L=256$,
$512$, and $1024$. It can be seen that the plots for BTW are not
parallel to each other. Especially at higher values of $q$, the plots
are dispersed and curved which do not allow to measure any critical
exponent. Whereas for the case of SSM various plots for different $q$
and $L$ values are parallel to each other for the whole range of
$\beta_{a}$ and the measured slope gives the value of
$\gamma_{sa}=1.784\pm0.004$. The values of other conditional critical
exponents are also measured following the same method and they are
found to be: $\gamma_{st}=1.547\pm0.004$, $\gamma_{at}=0.873\pm0.002$.
The exponent $\gamma_{xy}$ can also be obtained in terms of the
distribution exponents $\tau_x$ and $\tau_y$ as given in
\cite{santraPRE07}: $\gamma_{xy} =(\tau_y-1)/(\tau_x-1)$. This scaling
relation is satisfied within error bars for $x,y \in {s,a,t}$ for
SSM. For example, the values of $\tau_s=1.154$ and $\tau_t=1.236$
demand that $\gamma_{st}$ should be $1.532$, when the measured value
of $\gamma_{st}$ is found to be $1.547\pm0.004$. Thus, the extended
set of exponents obtained here for SSM from moment analysis of both
probability distribution and conditional expectation are consistent
with the scaling relations. It should be noted here that the exponents
$\tau_x$ and $\gamma_{xy}$ are found to be different from those
obtained for the same model on the regular and other fractal lattices.

\section{Time auto-correlation of toppling waves}
The multifractal scaling \cite{karmakarPRL05} in BTW model on regular
lattice is known to be due to the finite auto-corelation in the
toppling wave \cite{deMenechPRE00,deMenechPHYA02,stellaPHYA01}.  A
toppling wave is the number of topplings during the propagation of an
avalanche starting from a critical site without further toppling at
the same site and hence each toppling of the critical site creates a
new toppling wave \cite{priezzhevPRL96}. It is then importent to study
the time auto-correlation of the toppling waves for the BTW and SSM on
the percolation backbone. The time auto-correlation function is
defined as
\begin{figure}[t]
\centerline{\hfill
\psfig{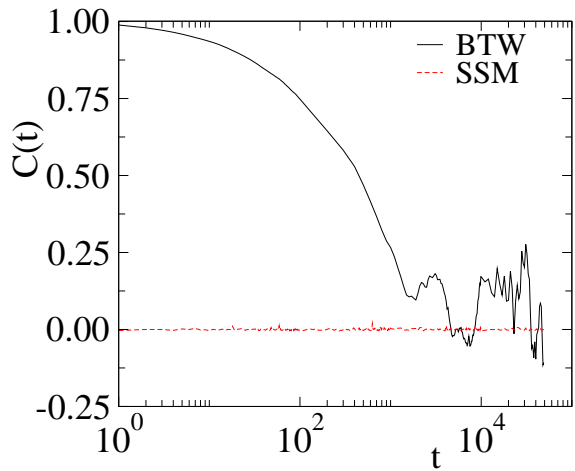} \hfill}
\caption{(Color online) Plot of auto-correlation function of toppling
  waves $C(t)$ against $t$ for BTW model (in solid black line) and for
  SSM (in red dashed line). For both the models $10^5$ toppling waves
  are considered and they are collected on one realization of the
  backbone for $L=1024$.}
\label{tauto}
\end{figure} 
\begin{equation}
\label{corl}
C(t) = \frac{\langle s_{k+t}s_k\rangle -  \langle s_k\rangle^2}{
  \langle s_k^2\rangle - \langle s_k\rangle^2},
\end{equation}
where $t=1,2,\cdots$ and $\langle\cdots\rangle$ represents the time
average. $C(t)$ is calculated for both the models --- on a system of
size $L=1024$, generating $10^{5}$ toppling waves in the steady
state. $C(t)$ values obtained are plotted against $t$ in
Fig. \ref{tauto}. It can be seen that $C(t)$ in the BTW is positive
and hence, the toppling waves are highly correlated, whereas for SSM,
the values of $C(t)$ is always $0$ revealing the uncorrelated toppling
waves. It should be emphasized here that Karmakar {\em et al.}
\cite{karmakarPRL05} showed that the toppling wave correlation in the
BTW-type sandpile model on an regular lattice is essentially due to
the precise toppling balance. Thus on the backbone the precise
toppling balance is maintained for BTW model and the toppling size
which consists of the correlated toppling wave, and the other
properties of avalanche like avalanche area and avalanche time do not
obey FSS rather than they obey multiscaling behaviour. The
uncorrelated toppling wave in SSM leads to the system to obey FSS
which is consistence with the observation in the multifractal analysis
in previous section.

\section{Toppling surface analysis}
\begin{figure}[t]
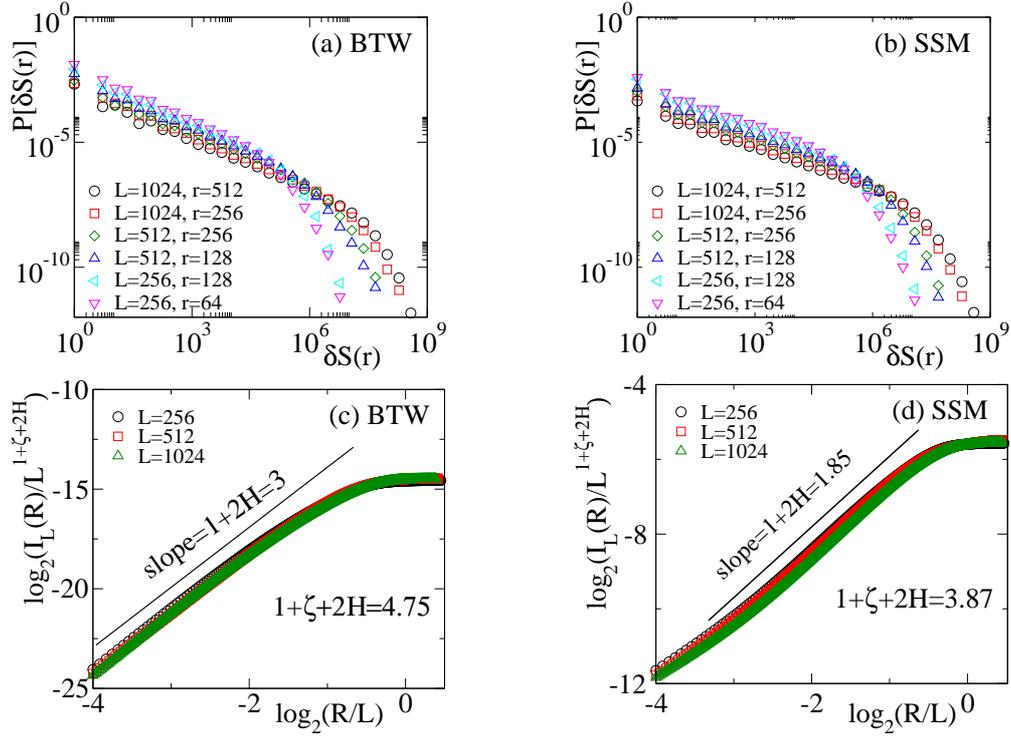

\centerline{\hfill 
  \psfig{file=DIST_delh_L256-1024_BTW.eps,width=0.35\textwidth}\hfill 
  \psfig{file=DIST_delh_L256-1024_SSM.eps,width=0.35\textwidth}\hfill
}
\centerline{\hfill 
  \psfig{file=IR_scaled_BTW.eps,width=0.35\textwidth}\hfill 
  \psfig{file=IR_scaled_SSM.eps,width=0.35\textwidth}\hfill
}
\caption{\label{pdfsr}(Colour online) Plot of probability distribution
  of $\delta S(r)$ for various values of $L$ and $r$ in (a) for BTW
  and in (b) for SSM. Plot of $I_{L}(R)/L^{1+\zeta+2H}$ against $R/L$
  in (c) for BTW and in (d) for SSM for
  $L=256(\Circle),512(\Box),1024(\triangle)$. The straight line in
  each figure having slope $1+2H$ is a guide to the eye.}
\end{figure}

The analysis of toppling surface of the avalanche, developed in
Refs. \cite{ahmedEPL10,bhaumikPRE14}, gives the deeper insight about
the topography of the avalanche structure. The values of the toppling
number of all the lattice sites of an avalanche define a surface
called toppling surface which is obtained for several large avalanches
whose area are more than $80\%$ of the total mass of the backbone on
which they occur. For a given system size $L$, a total of $N_{\rm
  span}=4096$ spanning avalanches are taken over $128$ different
configurations of backbone. The height of the toppling surface at a
position $i$ is given by $S(i)$, the toppling number at $i$th site on
the backbone. To study the scaling behaviour of toppling surface a
two-point height-height correlation function, the correlation between
the toppling numbers of two sites of backbone separated by a certain
distance is determined. The expectation value of the square of the
difference of toppling numbers $\delta S(r) =\left|S(x+r)
-S(x)\right|^2$ at two sites separated by a distance $r$ will give the
two-point height-height correlation function $C_L(r)$. To determine
the above said expectation the probability $P[\delta S(r)]$ of a
particular value of $\delta S(r)$ occurring for a fixed value of $r$
is estimated for several values of $L$ and $r$. Plots of $P[\delta
  S(r)]$ versus $\delta S(r)$ for various values of $L$ and $r$ are
given in Figs. \ref{pdfsr}(a) and \ref{pdfsr}(b) for BTW and SSM
respectively. Following Ref. \cite{bhaumikPRE14}, the form of the
probability distribution function $P[\delta S(r)]$ is proposed as
\begin{equation}
\label{psrpL}
P[\delta S(r)]=\frac{r^{-2H}}{L^\zeta} g\left[\frac{\delta
    S(r)}{L^{\zeta}r^{2H}}\right]
\end{equation} 
where $H$ is the Hurst exponent, $\zeta$ is another exponents, and $g$
is the scaling function. Thus for a given $L$, the correlation
function $C_{L}(r)$ is obtained as
\begin{eqnarray}
\label{crp}
C_{L}(r) &=&\int_0^\infty \delta S(r)P[\delta S(r)]
d[\delta S_{L}(r)] \nonumber\\ 
&\sim & r^{2H}L^{\zeta}
\end{eqnarray}
Note that $C_{L}(r)$ is a system size dependent correlation function
which is generally observed in stochastic sandpile models
\cite{bhaumikPRE14}. In order to determine the values of the Hurst
exponent $H$ and the other exponent $\zeta$, integrated correlation
function $I_{L}(R)$ up to a distance $R$ is obtained as
\begin{eqnarray}
\label{sicrf}
I_{L}(R) &=& \int_0^R C_{L}(r)dr \sim R^{1+2H}L^{\zeta},
\end{eqnarray}
It can be seen that at $R=L$ the value of $I_{L}(R)$ scales as
$I_{L}(L) \sim L^{1+2H+\zeta}$. Consequently a plot of
$I_{L}(R)/L^{1+2H+\zeta}$ against $R/L$ in log-log scale will give the
slope $1+2H$ and from the best collapse of data, one could find the
value of $\zeta$. The plots of $\log_2[I_{L}(R)/L^{1+2H+\zeta}]$
against $\log_2[R/L]$ for various values of $L$ are given in
Fig. \ref{pdfsr}(c) for BTW and in Fig. \ref{pdfsr}(d) for SSM. Tuning
the value of $1+2H+\zeta$ the best collapse is observed when
$1+2H+\zeta=4.75\pm 0.03$ for BTW and $1+2H+\zeta=3.87 \pm 0.01$ for
SSM; while the slope, which is equal to $1+2H$, is measured as $3.00
\pm 0.05$ and $1.85 \pm 0.02$ for BTW and SSM respectively (given by
the straight line in the respective figures). Thus the value of $H$ is
found to be $\approx 1$ and $0.425 \pm 0.010$ for BTW and SSM, while
the value of $\zeta$ is $\approx 1.75$ for BTW and $\approx 2$ for
SSM. It should be noted here that the value of the Hurst exponent $H$
ranges from $0$ to $1$ and its value defines the nature of correlation
presents in the surface, {\em e.g.}; the values $H > 1/2$, $H = 1/2$,
and $H < 1/2$ correspond to correlated, uncorrelated and
anti-correlated Brownian functions respectively \cite{meakin}. Since
the Hurst exponent of toppling surface of SSM studied on the backbone
is $0.42<1/2$, its toppling surfaces are anti-correlated surfaces. On
the other hand, the BTW toppling surface is expected to be smooth and
less fluctuating as usually seen when the model studied in the
two-dimensional square lattice. In-spite of the fact that the
substrate (percolation backbone) considered here is random as well as
fractal in nature, the toppling surface of the BTW model when studied
on such substrate is found to be completely correlated as the Hurst's
exponent is found as $\approx 1$.

\begin{figure}[t]
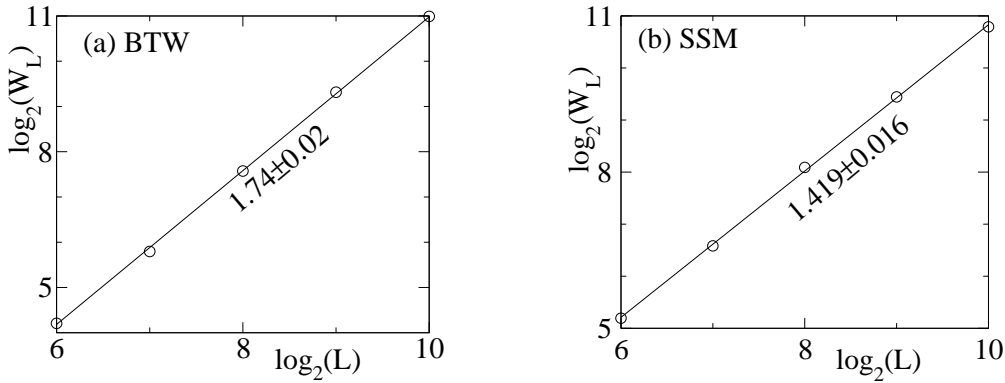

\centerline{\hfill 
  \psfig{file=Width_BTW.eps,width=0.35\textwidth}\hfill 
  \psfig{file=Width_SSM.eps,width=0.35\textwidth}
  \hfill}
\caption{\label{w_dcsr}(Colour online) Plot of width $W_L$ against $L$
  in double logarithmic scale for (a) BTW and (b) SSM. The slope of
  the solid straight line, obtained from the least square fitting, as
  indicated beside, is the estimated value of $\chi$ for the
  respective model. }
\end{figure}

To verify further the values of the exponents the overall surface
width $W_{L}$, for a given $L$ is also studied. $W_{L}$ is defined as
\begin{equation}
\label{w}
W_L=\left \langle \frac{1}{M_B} \sum_{i=0}^{M_B}(\bar S-S_i)^2
\right \rangle^{1/2}
\end{equation}
where $M_B$ is the mass of the backbone, $\bar S$ is the average
toppling height, and $\left \langle \cdots \right \rangle$ represents
the average over the toppling surfaces. The width $W_{L}$ is expected
to scale with $L$ as
\begin{equation}
\label{wl}
W_L \sim L^{\chi}
\end{equation}
where $\chi$ is known as the roughness exponent. To have an estimate
of the exponent $\chi$, $W_L$ is calculated for different system sizes
$L$ and plotted against $L$ in Fig. \ref{w_dcsr}(a) for BTW and
Fig. \ref{w_dcsr}(a) for SSM in double logarithmic scale. The
best-fitted straight line gives the slope as $\chi_{BTW}=1.74\pm0.02$
$\chi_{SSM}=1.419\pm0.016$. To obtain a relationship between the
exponents $H$ and $\chi$, the square of the width, $W_L^2$ can be
expressed as
\begin{eqnarray}
\label{wint}
W_L^2 &=& \frac{1}{M_B}\int_0^L C_{L}(r) rdr
\end{eqnarray}
As $M_B \sim L^{d_f^B}$ and the integration is over the plane of the
backbone, one could get 
\begin{equation}
\label{wint2}
W_L^2 \sim  L^{-d_f^B+\zeta+d_f^B+2H} \sim L^{\zeta + 2H} 
\end{equation}
which immediately follows a scaling relation
\begin{equation}
\label{czh}
\chi=\zeta/2+H. 
\end{equation}
This relation is well satisfied within the error bar by the measured
exponents. Note that this relation is also satisfied for SSM on $2$D
square lattice \cite{bhaumikPRE14} though the value of $\zeta$ equal
to $1$ there. The difference between Hurst exponent and Roughness
exponent appeared from system size dependent correlation function as
was observed in Ref. \cite{bhaumikPRE14}. The critical exponent of
toppling surfaces and that of the avalanche size capacity dimension
can be found to be related as
\begin{equation}
\label{Dschi}
D_s = \mbox{spatial dimension}+\chi = d_f^B+\chi.
\end{equation}
For the SSM, taking $d_f^B=1.64$ and $\chi_{SSM}=0.42$, $D_{s,SSM}$
should be $3.06$ which is in agreement to the measured value of
$D_{s,SSM}=3.062$ by the moment analysis of avalanche size. Since the
BTW model does not follow FSS, the value of $D_{s,BTW}$ is not well
defined and such relation is not satisfied well. However, taking
$d_f^B=1.64$ and the measured value of $\chi_{BTW}=1.74\pm 0.02$, one
can obtain $D_{s,BTW}=3.38$ close to the value of $\alpha_s(q)$ at
higher $q$ ({\em e.g} $\alpha_s(q=4)$ in Fig. \ref{falfa}(a)).

\begin{figure}[t]
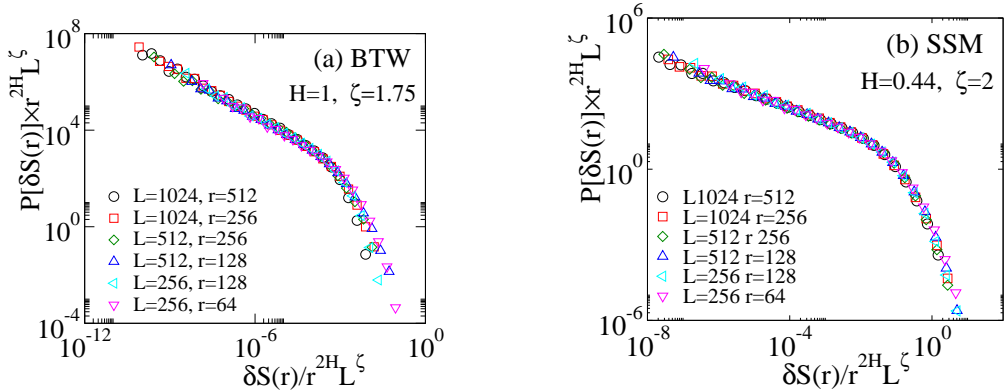

\centerline{\hfill 
  \psfig{file=DIST_delh_scaled_L256-1024_BTW.eps,width=0.35\textwidth}\hfill 
  \psfig{file=DIST_delh_scaled_L256-1024_SSM.eps,width=0.35\textwidth}
  \hfill}
\caption{\label{csr}(Colour online) Plot of Scaled distribution
  $P[\delta S(r)]r^{2H}L$ against scaled variable $\delta
  S(r)/Lr^{2H}$ for (a) BTW and (b) SSM for various choice of $L$ and
  $r$. Collapse of data for BTW model is not satisfactory whereas for
  SSM that is reasonably good.}
\end{figure}

Finally, to verify the scaling form of the probability distribution
$P[\delta S_{L}(r)]$, the value of the exponent $H$ and $\zeta$, a
scaled distribution $P[\delta S_{L}(r)]r^{2H}L$ against a scaled
variable $\delta S_{L}(r)/Lr^{2H}$ for different values $L$ and $r$
are plotted in Figs. \ref{csr}(c) and (d). Taking respective values of
$H$ and $\zeta$ of a given model, an attempt has been made to collapse
the data. It can be seen that while a good data collapse is observed
for SSM taking $H=0.42$ and $\zeta=2$, for BTW the collapsed data is
not satisfactory which could be due to the fact that BTW model does
not obey FSS ansatz. On the other hand, reasonable data collapse for
the SSM not only confirms the proposed scaling function given in
Eq. (\ref{psrpL}) is correct but also verify the measured correct
exponents.

\section{Conclusion}
Both the deterministic and the stochastic sandpile models have been
carried out on the percolation backbone in order to verify the effect
of a random fractal on the critical properties of such sandpile
models. By extensive numerical analysis, an extended set of critical
exponents of both the models has been estimated and verified through
various scaling analysis. Multifractal analysis of the probability
distribution functions and the expectations of the avalanche
properties suggest that though the spatial structure is a random
fractal, the BTW model preserves its multiscaling behaviour due to its
complete toppling balance, whereas the SSM retains its robust finite
size scaling behaviour. Moreover, the toppling surface analysis has
been carried out and the attempt has been made to explore the effect
of the fractal dimension of the backbone on the characteristics of the
toppling surface for both the models. New scaling relations have been
developed in terms of fractal dimension of the backbone and such
scaling relations are verified numerically. As the critical exponents
depend on the dimension of the underlying structures, the values of
the critical exponents of SSM on the percolation backbone are found to
be very different from those for the model defined on the regular and
other fractal lattices. Hence, SSM on the percolation backbone belongs
to a new stochastic universality class.

\bigskip
\noindent{\bf Acknowledgments:} This work is partially supported by
DST, Government of India through project
No. SR/S2/CMP-61/2008. Availability of computational facility,
``Newton HPC'' under DST-FIST project (No. SR/FST/PSII-020/2009)
Government of India, of Department of Physics, IIT Guwahati is
gratefully acknowledged.

\bibliographystyle{aipM}
\bibliography{ref}

\end{document}